\documentclass[
aps,%
12pt,%
final,%
notitlepage,%
oneside,%
onecolumn,%
nobibnotes,%
nofootinbib,%
noshowpacs,%
centertags]%
{revtex4}
\begin{document}
\selectlanguage{russian}

\noindent {\it ASTRONOMY REPORTS, 2013, 6}\bigskip \hrule  \vspace{15mm}

\title{AGN III~--- PRIMORDIAL ACTIVITY IN NUCLEI OF LATE-TYPE GALAXIES WITH PSEUDOBULGES}

\author{\firstname{\bf \copyright $\:$  2013 \quad  B.V.}~\surname{Komberg}}

\affiliation{Astro Space Center, P.N. Lebedev Physical Institute of Russian Academy of Sciences, Moscow, Russia}

\author{\firstname{\bf \copyright $\:$  2013 \quad  A.~A.}~\surname{Ermash}}

\affiliation{Astro Space Center, P.N. Lebedev Physical Institute of Russian Academy of Science, Moscow, Russia}

\begin{abstract}
\medskip\centerline{\footnotesize Received xx.xx.2012;$\;$ accepted xx.xx.2012}\bigskip\bigskip\noindent

1. Based on observational data on evolution of quasars and galaxies of different types along with the results of numerical simulations we make a conclusion that on low redshifts ($z<0.5$) QSOI/II objects in massive elliptical and spiral galaxies with classical bulges cannot be in late single activity event (be ``primordial''). 
Instead of it they have had events of activity earlier in their evolution.
It means that their presence on low redshifts is connected with the recurrence phenomenon, sequential wet minor mergings, because timescale of the activity does not exceed several units of $10^7$ years.

2. We define a new class~--- ``AGN III'' as active galactic nuclei in isolated late-type spirals with low-mass rapidly rotating pseudobulges.
We also state that only such objects can be in the primordial phase of activity at low redshifts. 
Black holes in such galaxies have masses $M_{BH}<10^7M_\odot$ and also, probably very high spin.
Such properties can explain their peculiar emission spectra.

A good representative of AGN III might be the galaxies with narrow (${\rm FWHM}(H\beta)\leq1200$ km/s) broad permitted emission lines~--- NLS.
It is believed that their black hole masses are less than $M_{BH}<10^7M_\odot$ and their host galaxies have pseudobulges instead of the classical ones.
Because host galaxies of NLS have pseudobulges and BLS (Broad-Line Seyfert galaxies) have classical bulges these two types of objects cannot have evolutionary connection. 
Presumably, the parent population of NLS are the quasars of ``population A'' \cite{dultzinmartinez2011}. 
\end{abstract}

\maketitle
\vspace{34mm}  \hrule  \vspace{4mm}
\noindent {\it\footnotesize E-mails of the authors:}  \qquad
\url{aermash@gmail.com,bkomberg@asc.rssi.ru}   

\newpage

\section{INTRODUCTION}

Since the discovery of the first quasar by M.~Schmidt during the half of a century it became clear that quasar phenomenon is a very active phase of evolution of galactic nuclei.
At present there are tens of thousands of known quasars with redshifts from $0.04$ to $\sim8$ and absolute magnitudes from $-23$ (by definition) to $-32$.
Only $\sim10\%$ are strong radio- or x-ray sources. About $90\%$ radiate only in optical or UV.
Space density of bright quasars grows rapidly from $z\approx0$ to
$z\approx2-3$ as $\rho_q\sim(1+z)^{10\pm1}$. For weak QSO the growth is less prominent ($\sim(1+z)^{2.3\pm0.7}$)
\cite{serjeantbertoldi2010}.

The quasar density has a maximum at $z\sim2-3$. At higher redshifts it slowly declines \cite{matteocroft2003}. 
The dependence $\rho_q(z)$ resembles the dependence of star formation rate on redshift \cite{schaererpello2007}.

Quasars like other AGNs are divided into two groups. Objects with wide permitted emission lines and low absorption are classified as QSO I. Objects with narrow permitted lines and high internal absorption~--- QSO II. This classification is an analogy of Sy1/Sy2 dichotomy\footnote{It should be mentioned that there are also many AGN of intermediate types.}.

Based of different estimates of the AGN activity timescale in quasar phase one can make a conclusion that it does not exceed $\sim10^7$ yr. (see \cite{komberg1971,schawinskievans2010}). 
From this it follows that quasars that started their activity at high $z$ cannot stay in single active phase till current epoch.
This is why I.S. Shklovsky asked his question about the nature of quasars at low redshifts: ``Are they either a result of the recurrence of the AGN phenomenon or of the recent strong flare of activity in galaxies which due to some reason delayed in their evolution?''\footnote{An attempt to answer these questions was made in \cite{komberg1984,komberg1989}}

\section{DIFFICULTIES OF THE HIPOTHESIS OF SIMPLE HIERARCHICAL CLUSTERING. ``THE DOWNSIZING''}

As became clear in recent years, there is no simple answer to the question about how do galaxies evolve. Character of the processes governing the evolution of star formation rate in different types of galaxies is very complex.
The numerical simulations and observations show that Hubble morphological classification establishes only to redshift $z\leq0.5$ (see \cite{feldmanncarollo2011,martigbournaud2012})). For more distant ($z>1$) galaxies this classification requires correction.

By now there are many observational facts contradicting to the classical hierarchic scheme of galaxy evolution, according to which masses and sizes of galaxies increase with time due to their mergings.

This contradiction was first discussed in \cite{cowie96}. Based on a sample of 280 starburst galaxies with ($z\approx0.8-1.6$) it was shown that their masses increase towards higher redshifts.

It was also shown that hierarchical clustering of dark matter halos, though able to explain formation of the Large-Scale Structure, does not imply the same evolution for barionic dissipative matter \cite{serrano2012}.
During recent years this conclusion finds more and more support.

For example, in \cite{merloni2009} it was shown that $M_{BH}\left/M_{\ast}\right.\sim(1+z)^{0.7}$, where $M_\ast$~--- the mass of stellar component.
In \cite{feldmanncarollo2011} authors stated that the process of formation of massive spheroidal galaxies is more rapid than formation of groups and clusters of galaxies.
At the same time in massive galaxies in massive DM-halos active star formation ends up earlier (to $z=1.5$) and they enter the regime of passive evolution.
Galaxies in less massive DM-halos continue forming stars up to $z=0$ \cite{poppingcaputi2012}.
The later the galaxy is formed, the younger its stellar population is with average age $\sim\left(\lg{M_{\ast}}\right)^{1.6}$ \cite{oserostriker2010}.
It should be noted that such evolution affects not only the stellar masses of galaxies but also the spins of their supermassive black holes.
Black holes in massive galaxies spin up earlier as a result of major mergings and lose their angular momentum later (at $z<2$) due to sequential minor mergings.

The problem of morphological evolution was addressed in many papers.
For example, in \cite{serrano2012} it was shown that S and Irr galaxies dominate in low-mass range at high z, while E/S0 galaxies dominate among the massive ones.
In \cite{delgadoserrano2010} authors studied two samples of galaxies $\Delta z_1=0.3-0.4$ and $\Delta z_2=1.3-1.5$ with $M_{AB}<-20.3$ selected from SDSS.
It turned out that in both ranges the fraction of elliptical galaxies is $\sim(3\div4)\%$ and the fraction of S0 is $\sim(13\div15)\%$.
At the same time the fraction of spiral galaxies increases from $\sim30\%$ in $\Delta z_2$ to $\sim70\%$ in $\Delta z_1$ while the fraction of irregular galaxies is $\sim10\%$ in $\Delta z_1$ and $\sim50\%$ in $\Delta z_2$.
From the above-mentioned the authors draw a conclusion that $\sim40\%$ of irregular galaxies turn into spiral galaxies with time, but not into E/S0 as one can expect from hierarchical clustering models.
Authors of \cite{vulcanipoggianti2011} came to the similar conclusion by comparing the abundances of different types of galaxies at $z=0$ and $z=0.8$:\\
\parbox{0.5in}{\ }The fraction of EG is $\sim40\%$ at both considered redshifts.\\
\parbox{0.5in}{\ }The fraction of S0 is $\sim13\%$ at $z=0.8$ and $\sim40\%$ at $z=0$.\\
\parbox{0.5in}{\ }The fraction of SG is $\sim40\%$ at $z=0.8$ and $\sim15\%$ at $z=0$.\\
The conclusion is that the transformation from SG to S0 occurs because the gas is removed from the galaxies near $z\approx0.4$.

It should be noted that there is no conventional point of view on the nature of S0 galaxies.
The authors of \cite{wilmanerwin2012} propose a hypothesis that there are two subpopulation of S0 galaxies~--- ``pre-processed'' and ``post-processed'' ones. 
The ``post-processed'' ones are the galaxies that were spiral in their past and lost their gas during the infall onto a cluster.
In \cite{elichemoral2012} it is stated that S0 galaxies can evolve along the Hubble sequence towards earlier types via dry minor and intermediate mergings.
But in \cite{silchenko2012} authors have the opposite point of view. They state that lenticular galaxies are the predecessors of spiral galaxies, which are the result of accretion of gas onto S0 galaxies followed by star formation.

We would like to mention that massive elliptical galaxies do not form a homogeneous population. Instead they are divided in two groups based on the light profile~--- the core EG and the disky EG (see, for example, \cite{hamilton2010}). They can be the hosts of radio-loud QSO and radio-quiet QSO, respectively. 
The observations showed that stellar mass $M_{\ast}$, above which EG become dominant, increases from $z=0$ towards $z=2$. This is what is actually called ``downsizing''. 
We also would like to mention that observations of massive EG by \cite{hopkinsbundy2007} confirmed the theoretical prediction that QSO phase, connected with wet mergings (the blue sequence) stimulates warming up and removing of gas in the host galaxy. Which in turn leads to quenching of star formation and transition of the galaxy to the red sequence.

All massive core EG have effective radii $r_e>10$ kpc, $P_r>10^{22.5}$ W/hz and X-ray emitting coronas. Less massive disky EG have $r_e<10$ kpc, $P_r<10^{21.5}$ W/hz.
In \cite{huertascompany2010} authors hypothesized that red core EG are formed in dry major mergings, while blue disky EG are formed via wet minor mergings. It explains their rapid growth in size as $\sim M^2_\ast$.

It should be admitted that it is not yet clear how the core EG and disky EG are formed.

One cannot of course write off other evolutionary mechanisms. Their traces can be detected by properties of the circumnuclear disk, presence of binary black holes etc.
Future numerical simulations can help to answer which types of galaxies are formed via mergings of galaxies of different types.
For example, will the systems formed via mergings of E+S, S0 + S, E + S0, S + S with different fractions of gas differ and how \cite{scarlata2007}.

\textbf{Conclusion:}
All above-mentioned implies that galaxy evolution depends on many factors, such as initial conditions and properties of their close surroundings and the place in the Large-Scale structure.
That is why it is not yet possible to determine morphological types of today's galaxies predecessors.

In the recent years there were some publications presenting evolutionary schemes describing the ``downsizing'' phenomenon.
For example, in \cite{oserostriker2010,johansson2012} a model with two phases of star formation was proposed:\\
\underline{An early, rapid stage at $z>3$ (in situ).} Monolithic collapse of intergalactic gas leads to formation of the inner (<3kpc) part of the galaxy, where stars are formed.\\
\underline{A late stage, at $z<3$ (ex situ).} This phase is more prolonged. Accretion of low-mass stellar complexes with old stellar population (``dry minor mergings'') leads to formation of peripheral parts of a galaxy
The role of this phase increases with increase of $M_\ast$ and decrease of $z$.
According to this two-phase scheme, more massive galaxies contain larger fraction of old stars in their periphery. The relation of amount of stars formed in early and late phases depends on the mass of the galaxy (see \cite{greenemurphy2012}).

\textit{
Due to the fact that quasars are active nuclei in massive galaxies which, because of the above-mentioned, are formed in early epochs ($z>1.5$) it is becoming clear that nearby quasars ($z<0.5$) are the result of recurrent reactivation in nuclei of old galaxies.
It can be caused by injection of accreting matter during wet minor mergings.
The connection between mergings of galaxies and nuclear activity can be traced by the dependence of a relative frequency of AGN in close pairs on the redshift and distance between components.
}

\section{THE CONNECTION BETWEEN GALAXIES AND PROPERTIES OF THEIR CENTRAL REGIONS: BULGES AND PSEUDOBULGES.}

For example, in \cite{kossmushotzky2012} it was shown that there are only $1\%$ of Sy1 galaxies in pairs with separation less than $30$ kpc.
The fraction of Sy1 reaches $15\%$ when separation is less than $15$ kpc. 
Some traces of past mergings are observed in $40\%$ of Seyfert galaxies \cite{smirnovamoiseev2010} and the fraction of galaxies in pairs increases with redshift as $z\sim(1+z)^{0.6\pm0.5}$ \cite{manzirm2011}.
In \cite{buitragotrujillo2013}, using a sample of 1100 massive galaxies with redshift $z<3$ it was shown that there are many disk-like galaxies at $z>1$, but towards lower redshifts the fraction of ellipticals reaches $70\%$.
In \cite{carpinetikaviraj2012} authors analyzed a sample of 3373 disk galaxies.
They drew a conclusion that nuclear activity is observed only on late phases of merging significantly after a  starburst.
Among galaxies in the ``post-merging'' state there are $\sim90\%$ of SyG and LINERs, about $16\%$ of starburst galaxies and the same amount of quiescent galaxies.

From all above-mentioned it is clear that merging/interaction between galaxies, if at least one of them is gas-rich leads to a delayed burst of nuclear activity. 
The process of transferring gas towards galactic nucleus is complex and takes significant amount of time (see, for example, \cite{alexanderahickoxa2012}).
\textit{But the simulations showed that one can derive previous merging history judging on properties of galactic disks.
Even if galaxies are single in the present time and show overall regular morphology.}
The parameters of the so-called ``classical bulges'' (the bulge mass, luminosity and stellar velocity dispersion $M_{bulge}$,$L_{bulge}$,$\sigma_\ast$) correlate with masses of supermassive Black Holes (see, for example, \cite{gadotti2008}).
Numerical simulations did show (for example, \cite{wandel2011,martigbournaud2012}) that classical bulges with old stellar population typical for massive galaxies are formed in major mergings at early epochs.
But in galaxies formed in later epochs bulges differ from the classical ones in their properties.
They are less massive, they have higher aspect ratio and, on average, a younger stellar population.
Such bulges are called ``pseudobulges'' (for example, \cite{kormendykennicutt2004}).
The classical bulges resemble spherical stellar systems, the pseudobulges show instead properties similar to disk-like systems and their stellar population is, on average, younger (see, for example, \cite{shankar2012}).
It is accepted that the threshold value, delimiting pseudobulges and the classical ones is $B/T\leq0.2$ and the Sersic index\footnote{$\mu(r)=\mu_0+b_n\left(\dfrac{r}{r_e}\right)^{\frac{1}{n}}$, where $\mu$~--- surface brightness, $b_n=0.87n-0.14$. De-Vaucouleur profile has $n=4$, $n=1$ corresponds to an exponential disk.}
$n<2$ for pseudobulges (for example, \cite{xivrydavies2011}).
It should be noted that distributions of the Sersic indexes for the classical and pseudobulges overlap.

It is believed that in contrast with the classical bulges the pseudobulges form via processes of internal secular evolution caused by disk instabilities. Their host galaxies did not experience major mergings at $z<1.5$.
Different nature of the pseudbulges is also evidenced by the fact that they have significantly smaller $M_{BH}$ on the $M_{BH}$--$M_{bulge}$ plane (see, for example, \cite{wandel2011}).

Some authors came to the conclusion that differences between the classical and pseudobulges (see, for example, \cite{gadotti2008}) are not due to the differences in their stellar masses derived from $\sigma_\ast$, but due to differences in angular momentum. The pseudobulges are rotationally supported whereas classical bulges are not. 
\textit{It is possible that this fact can stand for higher spins of black holes in pseudobulges. This will definitely affect the mechanisms of AGN activity.}
It should be stressed that there is no ultimate answer on the connection between black hole masses and their spins. It depends on the mechanisms of evolution of galaxies and their bulges (for example,
\cite{sikora2009,cervantessodi2011}).

In many papers results contradicting with the standart model of NLS objects are presented.
For example, according to \cite{doinagira2012}, some of these objects have extended radio structures, which were observed on VLA.
Authors estimated that the energy required for their formation is of order $\geq10^{44}$ erg/s.
Such luminosity seems to be to large even for NLS accreting at the Eddington rate with black hole mass $M_{BH}\leq10^7M_\odot$.
Due to this, one should consider additional mechanisms of energy emission in central engines of NLS objects.
The question about the influence of the black hole rotation on properties of the central engine is still open.
For one of the NLS (NLS IRAS13224-3809) with rapid x-ray variability ($\Delta t\sim 100$ sec) authors of \cite{fabiankara2012} estimated black hole spin as $a>0.98$ \footnote{$a$~--- dimensionless parameter characterizing the spin of black hole (Kerr parameter) $a\equiv\dfrac{cJ}{G M_{BH}^2}$ (see, for example, \cite{wangmao2011})}.
Different mechanisms of energy extraction from a rotating black hole were discussed, such as the Blanford-Znajek \cite{blandfordznajek1977} or Penrose proccesses in the presence of magnetic fields \cite{dadhich2012}.

It should be noted that in the recent years massive BHs were also discovered in bulgeless galaxies (see, for example \cite{satyapal2008}).
It means that a mechanism of black hole growth might be connected not only with bulge parameters, but also with parameters of dark matter halo.
In \cite{simmons2012} it is said that in bulgeless host galaxies of AGN exists the following relation: 
$M_{BH}$--$M_\ast^{tot}$; see also \cite{marleau2012}.
This conclusion is supported by the fact that there is a fair correlation between $M_{BH}$ and opening angle of spiral arms (for example, \cite{seigar08},\cite{treuthardt2012}), which, in turn, correlate with global properties of galaxies.
It is not yet clear how do pseudobulges really form.
For example, in \cite{keselman2012} it was shown that in the $\Lambda$CDM-simulations pseudobulges can form through mergings of gas-rich spiral galaxies with weak disks. The stability of such bulges is enforced by their rapid rotation.
In the same work in was shown that massive disk galaxies can form through mergings of gas-rich galaxies with pseudobulges or even bulgeless ones.
It would be interesting to study the distribution of their black hole masses.

\textbf{Conclusion:}
The above-mentioned facts do not contradict with the hypothesis that not too massive late-type spiral galaxies with pseudobulges during their evolution since $z\sim2$ did not experience significant mergings because they have pseudobulges and their black holes are actually small.  
And this in turn can lead to the observed peculiarities in their AGN emission.

\section{THIRD TYPE OF ACTIVE GALACTIC NUCLEI~--- AGN III}

We have already mentioned that there are two types of AGN.
This classification is based on the line widths.
Objects with broad permitted lines are AGN I, the ones without broad permitted lines and with high obscuration are classified as AGN II.
Objects with strong variability of non-thermal continuum sometimes show transitions from one type of AGN to another. When continuum significantly decreases AGN I can turn into AGN II and vise versa. Such events were observed in some Seyfert galaxies.
This probably happens because the distance from the ionizing continuum to region where broad lines form depends on intensity of ionizing continuum.
These conclusions strongly depend on assumptions on geometry of line-emitting region.
It is either an accretion disk or a gaseous shell surrounding radio jets.
It also must be taken into account that AGN II themselves can be divided in two subtypes.
The genuine AGN II with no broad lines present and obscured-AGN II with a high internal obscuration by dusty torus.
In the latter AGN broad lines sometimes can be seen in a polarized light.
The obscured AGN II are by actually AGN I.

We think that there is enough observational evidence to define a new class of AGN~--- AGN III besides existing AGN I and AGN II.
The point is that host galaxies of AGN I and AGN II are massive spheroidal systems (in case of quasars and radiogalaxies) or massive disk systems (in case of RQ QSO or Seyfert galaxies) with classical bulges. Their black hole masses exceed $M_{BH}>10^7M_\odot$.
In the Local Universe the stellar mass is distributed among different types of galaxies in the following way (\cite{gadotti2008}): elliptical~--- $35\%$, spiral galaxies~---$36\%$; galaxies with classical bulges~--- $25\%$, with pseudobulges~--- $3\%$, with bars~--- $4\%$.
In contrast with AGN I and AGN II, the proposed AGN III group consists of isolated late-type galaxies with pseudobulges and black holes masses $M_{BH}<10^7M_\odot$.
They also have some other peculiar properties, such as relatively narrow broad permitted emission lines (FWHM$\leq2000$ km/s).
There is growing interest in such objects during the recent years.

In \cite{woltergioia2005} 3 objects with x-ray luminosity $L_{XR}\sim10^{44}$ erg/s and very narrow lines $\Delta V_{\frac{1}{2}}({\rm H}\alpha,{\rm H}\beta)<750$ km/s were studied. Due to their properties they can be classified as QSOII.
At the same time they do not show significant obscuration in soft x-ray. It means that their broad line regions are not obscured from the observer.
According to \cite{stern2012}, a search of AGN II objects in low redshift samples is simplified by the presence of strong correlation between x-ray luminosity $L_{2kev}$ and $L_{{\rm H}\alpha}$.
In such sample one can select objects with line widths less than $2000$ km/s and without absorption in soft x-ray.
These will be the above-mentioned AGN III objects.

Let us return to the connection between AGN III activity and the properties of the bulge.

For example, in \cite{jiang2011} authors studied two samples of galaxies selected from HST data with $M_I<-19$ and $M_{BH}<10^6M_\odot$.
The first one included 173 objects without any traces of activity in their nuclei. More than $90\%$ of their host are extended disk galaxies.
The second one included 147 galaxies ($z<0.35$) with active nuclei. Only $7\%$ of them have close companions. 
The authors came to conclusion that in both samples properties of bulges are similar and they can be classified as pseudobulges.
This leads to conclusion that active nuclei can be fed by secular processes in pseudobulges, which means accretion of cold gas driven by instabilities in galactic disk.

Based on the fact that formation of pseudobulges does not involve major mergings, authors came to the conclusion that black holes in this galaxies are primordial to their host galaxies.
All this facts imply that the formation processes of classical and pseudobulges have different nature (\cite{serrano2012,xivrydavies2011,mathur2012}).
But in centers of such bulges supermassive black holes can be formed and accrete matter, what leads to nuclear activity.
It should be noted that in some of disk galaxies the Nuclear Star Clusters~(NSC) are observed besides the supermassive black holes.
According to \cite{scottgraham2012} in late-type spirals with dynamical mass of the spherical component $M_{sph,dyn}>5\times10^9M_\odot$ mainly black holes are formed.
But if the mass of the spherical components is in the range $10^8$--$10^{10}M_\odot$, supermassive black holes and nuclear star clusters can be present simultaneously.
But according to \cite{antonini2012}, NSC are not observed in galaxies with an absolute magnitude less than $M_B=-12^m$.
Generally speaking, NSC are bluer than the host galaxy and they also rotate rapidly.
It is possible that NLS can be formed through mergins of globular clusters that lose their angular momentum by dynamic friction and fall onto the galactic center.
It is quite interesting that among Seyfert galaxies objects with $M_{BH}<10^6M_\odot$ can be found.
For example, in \cite{xiaobarth2011} for a sample of 76 Sy1 there was obtained a relation 
$M_{BH}(\sigma_\ast)$
and it was also shown than it does not differ from the relation:
\[\lg{M_{BH}}=(7.68\pm0.08)+(3.32\pm0.22)\frac{\sigma_\ast}{\rm200 km/s}\]
for Seyfert galaxies with black hole masses $M_{BH}>10^6M_\odot$.
However, as we have already mentioned above, the $M_{BH}$--$M_{bulge}$ relations for BLS and NLS actually differ.

\textit{
Finally, as we showed above, we can say that the answer to the question if quasars in the Local Universe are young is most likely negative.
But if we consider much fainter objects such as active nuclei in local disk galaxies (AGN III), the answer to the question if they are primordial might be positive.
Their host galaxies are low-mass and formed later. The presence of pseudobulges and low-mass black holes testify to the absence of major mergings in their evolution. Which in turn leads to the conclusion that the activity in such objects cannot be recurrent.
}

\section{Narrow-line Seyfert 1 galaxies}

Let us consider typical representatives of AGN III class~--- the NLS.
They have broad permitted lines much narrower that in the classical Seyfert 1s.
NLS were first identified in \cite{osterbrock1985} as Seyfert 1 galaxies with widths of broad permitted emission lines less than ${\rm FWHM}({\rm H}\beta)\leq2000$ km/s.
Their spectra show some more peculiarities beside narrow H$\beta$ lines. They have strong FeII emission and weak [OIII] $\lambda$5007 \AA~\cite{komossa2008}.
But all these criterions need to be corrected for luminosity.
Despite the fact that NLSy1 might not be a uniform population, all of them have pseudobulges, black hole masses less than $M_{BH}<10^7M_\odot$ and high Eddington ratios (see review \cite{kombergermash2011}).
There are less than 10\% of NLS in optical samples. But they constitute $\sim15\%$ of hard x-ray AGN samples and up to 30\% of soft x-ray AGN samples \cite{dultzinmartinez2011,castellomor2012}.
It should be stressed that optical and x-ray NLS samples are not identical.
About 5\% NLS are radio-loud and have SEDs similar to BLAZARs.
Some of them have even significant $\gamma$-ray fluxes (see \cite{yuanzhou2008,foschini2011}).

We have mentioned earlier that some objects with high luminosities usually classified as QSO II resemble in some properties the NLS, especially if its host galaxy is a spiral one. 
This issue was discussed in \cite{dultzinmartinez2011} (see also more recent work by the same authors \cite{marzianisulentic2012}).
There are $\sim30\%$ of objects with such properties ($\Delta V_{1/2}=2000-4000$ ª¬/á) in the Palomar-Green sample of bright quasars \cite{borosongreen1992}.
Authors classified them as Pop. A. Part of this population with $\Delta V_{1/2}\leq2000$ km/s are NLS galaxies. 
On the $\Delta V_{1/2}({\rm H}\beta)$--$L_{bol}$ plane the boundary between NLS and Pop. A objects is proportional to $L_{bol}^{0.67}$

It is possible that objects with properties of QSO POP A may evolve into NLS1.
Such a hypothesis can be proposed if properties of the center of our Galaxy is considered.

There are some characteristics of the center of the Milky Way that allow it to be classified as AGN III \cite{hammerpuech2012,capuzzodolcetta2011}. 
Our closest neighbor, M31 (The Andromeda Galaxy) is a regular quiescent disk galaxy with the Hubble type Sbc, the black hole mass $M_{BH}=3\times10^7M_\odot$. In contrast, our Milky Way has a rare type. Such galaxies constitute only 1\% of all disk galaxies. It has a pseudobulge, black hole with mass $M_{BH}=4\times10^6M_\odot$ and an active nucleus (Sgr A).
The radio and x-ray fluxes of Sgr A are $\sim10^{35}$ erg/s and $L_{XR}\geq10^{37}$ erg/s respectively. 
To explain this unusual characteristics the authors make a presumption that Milky Way did not experience wet major mergings during recent $\sim10^{10}$ years, but only dry minor mergings.

This is why the nucleus of our galaxy has evolved via accretion of cold gas from the disk.
We cannot exclude the possibility that in the past our galaxy has had an active QSO POP A nucleus.
The remnants of this phase of activity are observed now as extended $\gamma$-emitting regions near the center of our galaxy \cite{guomathews2012,zubovas2011}.

We would like to stress that in the previous papers by \cite{marziani2001,boroson2002} authors built a relation $\Delta V_{1/2}({\rm H}\beta)$-- $W_{FeII\lambda4570}/W_{{\rm H}\beta}$ or PC1--PC2 in terms of principal components for different types of AGN.
This relation for AGN is an analogue of the Hertzsprung-Russell diagram for stars.

Such a similarity between stars and AGN exists because accretion disks are optically thick and emit quasithermal continuum from UV to IR. Thus, in some sense they resemble stars despite the fact that energy sources are of completely different origin.

Local neighborhood of NLS galaxies does not differ from the neighborhood of galaxies of the same luminosities and morphological types.
It is clear that they are absent in dense areas and in voids.
A typical location of NLS is in filaments and periphery of clusters (see \cite{xukomossa2012}). 

In some papers (for example, \cite{bian2003}) a conclusion was made that NLS galaxies evolve into BLS by the bulge and black hole growth. 
Nevertheless, authors of other works (like \cite{nerilarios2011}) argue that such a transformation is impossible because it would require a merging with a gas-rich galaxy.
The observations showed that NLS are isolated galaxies and they are usually not in pairs, even not in close ones.

Of course, a slow evolution towards growth of bulge and black hole masses can occur even without wet major mergings. 
But this process is slow and its timescale is larger than the evolution timescale of an active nucleus.
But, some authors (for example, \cite{bian2003}) share the point of view that classical bulges can evolve into pseudobulges via minor mergings on timescale of $\tau\backsimeq10^8$ years.

We would like to stress that the classification of quasar candidates into QSO I, QSOII and QSO III is not an ultimate one. Different sets of types can be defined using another sets of parameters for classification.
This situation is similar to elaboration of intermediate subtypes between Sy1 and Sy2.
For example, in \cite{steinhardt2011} authors selected quite large (about $20\%$) group of objects from the SDSS QSO sample.
They called these objects ``Anomalous Narrow-Line Quasars''.
In their spectra the narrow component of the H$\beta$ line has FWHM larger than 1200 km/s and correlates with FWHM of the broad component of the H$\beta$ line.
The authors explain this peculiarity in the ANL spectra by the influence of strong circumnuclear wind, reaching the narrow-line region and leading to the widening of the lines.

\section{CONCLUSION}

We should consider here new theoretical and observational aspects involving AGN III objects and, in particularly, NLS.

1. Due to the fact that $\sim7\%$ of NLS are radio-emitters and their SEDs are similar to those of BLAZARs, they can be of interest for observations on radio interferometers and also in x-rays and gamma rays.
Though their fluxes are not as high as those of BLASARs, because of their relative proximity they pose some interest, especially if their variablity is taken into account.
Moreover, additional features related with a relativistic speed of rotation of their supermassive black holes may be observed.

2. There is an intriguing prospect of studying AGN III in ${\rm H}_2{\rm O}$ (1.35 á¬) maser emission in order to find out the exact emitting regions and the pumping mechanism.
According to \cite{constantin2012,tarchicastangia2011} there is an unexpectedly high detection rate of ${\rm H}_2{\rm O}$ masers in NLSy1.
It contradicts with the commonly accepted picture that maser lines are formed in circumnuclear disks and are typical for Sy2 and, therefore, should not be present in Sy1 population.
It seems that water masers in NLS are formed in outflows or somehow connected with the jets.

3. The question about globular clusters (GC) in these objects is still open.
It seems that the expected amount of GCs in AGN III is quite low. There was discovered a relation between the amount of GCs and masses of central black holes in host galaxies (see \cite{burkert2010}). 
In other types of spiral galaxies some of the black holes coexist with nuclear star clusters (\cite{marelrossa2006}).
This question is still unanswered for NLS and requires a close consideration.

4. From the assumption that the evolution of galaxies with AGN III nuclei is not related with mergings some observational predictions might follow.

4.1 NLS-like objects should not be observed at redshifts higher than $z>1-2$ because the timescale of pseudobulge formation is large.

4.2 Binary black holes should not be present in AGN III nuclei because they did not experience major mergings.

4.3 If the predecessors of BLS are the RQ QSO1, it is tenable to suppose that QSO POP A(\cite{dultzinmartinez2011}) objects might be the predecessors of NLS.
The fact that luminosity functions of these object merge at the bright end \cite{ermash2012} supports this hypothesis. 

The evolution of NLS into BLS seems less likely because it implies evolution of a pseudobulge into a classical one. This process requires major mergings.
It should be noted that in some papers there were proposed other mechanisms of bulge growth in disk galaxies, see \cite{mathur2012}.

4.4. Our assumption made in 5.3 leads to conclusion that in the past there could be a much stronger activity in NLS nuclei (QSO POP A).
Remnants of this activity may still show themselves in the current epoch, similar to the $\gamma$-ray bubbles in our galaxy.
A future discovery of such features in NLS galaxies in gamma rays may serve as a confirmation for our proposed evolutionary mechanism: QSO POP A => NLS.

\medskip
The authors would like to acknowledge A.V.Zasov for interesting discussions and useful comments.

This work was supported by the Basic Research Program of the Presidium of the Russian Academy of Sciences ``Nonstationary Processes in Objects in the Universe''(P-21) and 
by The Ministry of education and science of Russian Federation, project 8405
and The Program of state support of Leading Scientific Schools in Russian Federation (grant SC-2915.2012.2 ``The formation of the Large-Scale structure of the Universe and cosmological processes'').

\bibliography{paper_agniii}  
\end{document}